\documentclass[conference]{IEEEtran}
\IEEEoverridecommandlockouts

\usepackage{cite}
\usepackage{amsmath,amssymb,amsfonts}
\usepackage{algorithmic}
\usepackage{graphicx}
\usepackage{textcomp}
\usepackage{xcolor}
\usepackage{comment}
\usepackage{hyperref}
\def\BibTeX{{\rm B\kern-.05em{\sc i\kern-.025em b}\kern-.08em
    T\kern-.1667em\lower.7ex\hbox{E}\kern-.125emX}}
\begin{document}

\title{Using Psychophysiological Insights to Evaluate the Impact of Loot Boxes on Arousal
}

\author{\IEEEauthorblockN{1\textsuperscript{st} Gianmarco Tedeschi}
\IEEEauthorblockA{
\textit{IT University of Copenhagen}\\
Copenhagen, Denmark \\
gite@itu.dk}
\and
\IEEEauthorblockN{2\textsuperscript{nd} Rune Kristian Lundedal Nielsen}
\IEEEauthorblockA{\textit{Play, Culture and AI Section} \\
\textit{IT University of Copenhagen}\\
Copenhagen, Denmark \\
rklb@itu.dk - 0000-0002-7209-9041}
\and
\IEEEauthorblockN{3\textsuperscript{rd} Paolo Burelli}
\IEEEauthorblockA{\textit{Play, Culture and AI Section} \\
\textit{IT University of Copenhagen}\\
Copenhagen, Denmark \\
pabu@itu.dk - 0000-0003-2804-9028}
}

\maketitle

\begin{abstract}
This study investigates the psychophysiological effects of loot box interactions in video games and their potential similarities to those recorded during gambling interactions. Using electrodermal activity (EDA) measurements, the research examines player arousal during loot box interactions and explores the relationship between Internet Gaming Disorder (IGD) severity and loot box interactions from a psychophysiological perspective.
The study employs a custom-designed game to control experimental conditions and standardise loot box interactions. Participants' IGD severity is assessed using the Internet Gaming Disorder Scale – Short Form (IGDS9-SF), while arousal is measured through EDA, analysing both tonic and phasic components. The study contributes to the ongoing debate surrounding gaming disorder and loot boxes, offering insights for game developers and policymakers on the potential risks associated with random reward mechanisms in video games.
\end{abstract}


\section{Introduction}
The recognition of gaming disorder as a mental health condition has been a pivotal moment in the ongoing discourse on problematic gaming. The American Psychiatric Association introduced Internet Gaming Disorder (IGD) as a condition warranting further study in the DSM-5 \cite{b1}, followed by the World Health Organization (WHO), which included gaming disorder in the ICD-11 \cite{b2}. Both classifications identify persistent and problematic gaming behaviours that negatively impact an individual’s personal, social, or professional life. The Internet Gaming Disorder Scale – Short Form (IGDS9-SF), widely adopted in research, was created to evaluate the nine traits' severity as outlined in the DSM-5 \cite{b3, b4}. However, these classifications sparked a debate among scholars who were divided on the research base, the symptomatology, and the comparison to substance use and gambling disorder (GD), with the critics arguing that the comparison lacks specificity and leans on pathologizing normal behaviours in non-problematic video game players  \cite{b5, b6, b7, b8}.
The growing implementation in games of gambling-like random reward mechanisms (RRMs) reinforces the conceptual overlap between gaming and gambling \cite{b9}, and has been at the centre of scrutiny due to their structural \cite{b10} and psychological \cite{b11} similarities to gambling. While loot boxes may not always involve financial risk, their presentation often resembles gambling, employing flashing, dramatic, and suspenseful visual and auditory cues that mimic the sensory experience of slot machines \cite{b12}. Researchers focused on loot boxes’ potential to encourage addictive spending patterns, particularly among younger players \cite{b13}, suggesting that exposure to loot boxes at an early age may contribute to the development of gambling behaviours later in life \cite{b14}. Moreover, researchers consistently highlight similarities in impulsivity and reward-seeking behaviours between loot box interaction and gambling \cite{b15} as well as a moderate correlation between loot box spending and both problem gambling and excessive gaming \cite{b16}.
Player engagement in video games is a complex phenomenon, and several questionnaires have been developed to evaluate player experience, but self-reported data are subject to recall biases, prompting researchers to adopt psychophysiological methods as objective alternatives \cite{b17}. Electrodermal activity (EDA), used to evaluate stress \cite{b18} and arousal \cite{b19}, electromyography (EMG), used to evaluate emotional expression \cite{b20}, electroencephalography (EEG), used to evaluate cognitive engagement, \cite{b21}, and electrocardiography (ECG), used to investigate stress and frustration \cite{b22}, are some of the psychophysiological tools with many possible applications in game user research. Studies on gambling addiction have extensively utilised psychophysiological tools \cite{b23}.

\subsection{Research Goal}
This study builds on previous findings to investigate the impact of loot box interactions on players' physiological responses, specifically assessing whether these interactions trigger measurable changes in player arousal and whether people with a higher IGD severity show hyposensitivity, similar to patterns observed in problematic gamblers \cite{b24}. To the authors’ knowledge, only a few attempts have been made to investigate whether the psychophysiological responses observed in gambling addiction studies also manifest during loot box interactions \cite{b25}. By addressing the current lack of physiological evidence in loot box research, the study contributes to the debate on the symptomatology of the gaming disorder and to the comparison between RRMs and gambling. The results have implications not only for academic debates but also for game designers and regulators.

\section{Methodology}
\subsection{A Minute Outside}
For this study, we used \textit{A Minute Outside} (2025)\footnote{https://ab0-22.itch.io/a-minute-outside}, a 2D top-down shooter survival game redesigned to meet the research requirements, which allowed for a high degree of experimental control, ensuring consistency in session length and standardising loot box interactions. This approach also allowed the manipulation of randomised rewards, which would not have been feasible with an existing commercial game.
The core gameplay loop involves exploring a post-apocalyptic wasteland, fighting off endless waves of enemies, and collecting resources within a strict time limit. The players can go back to their base to use the retrieved resources to purchase a loot box. These loot box interactions follow a standardised reward order to ensure that all participants experience the same sequence of weapon acquisition. More information is available on the video game website.

\subsection{Operationalisation of Constructs}
IGD was scored across participants through the IGDS9-SF. Each item was rated on a five-point Likert scale (1 = Never; 5 = Very Often), with higher scores indicating greater IGD severity. A trait is considered severe if the related item received a score equal to or greater than 4 (Often), meaning that the threshold used to determine whether the IGD is present in the participant is set to 36 \cite{b4}.
Arousal was measured via exosomatic direct current (DC) recording of EDA, and three parameters were extracted: skin conductance responses (SCRs) and Area Under the Curve (AUC) derived from the phasic component, and skin conductance level (SCL) from the amplitude of the tonic components \cite{b26}. SCR amplitude and AUC were analysed to examine players' responses to loot box interactions, similar to \cite{b25}, where researchers found that individuals with GD exhibit hyposensitivity to wins, which translates into shorter SCRs ranges, suggesting a diminished reward response that differs from non-problematic gamblers.


\subsection{Participants and Protocol}
The experiment was conducted in an electromagnetically isolated booth to minimise noise and signal interference. Physiological data were collected using the Shimmer3 GSR+ device for EDA, which was recorded at a sampling rate of 100 Hz and transmitted data via Bluetooth to the ECL Shimmer Capture interface. The device electrodes were placed on the proximal phalanx of the thumb and little finger of the mouse hand to minimise movement-related noise, considering that the other fingers were used for gameplay input.
The study recruited 13 participants, but only 11 (\textit{age}, \textit{M} = 27.63; \textit{male}, n = 9, 81.8\%; \textit{female}, n = 2, 18.2\%, \textit{other}, n = 0) provided valid data. The inclusion criterion was familiarity with mouse-and-keyboard controls to ensure that all participants engaged with gameplay naturally. The participants provided their informed consent and completed a preliminary survey, collecting data on age and weekly gaming hours, and answered the IGDS9-SF (M = 18.72, SD = 6.89, SE = 2.08, Cronbach's alpha = 81\%). The participants' scores ranged from 10 to 28, meaning that none of them had significant IGD. Weekly gaming hours are included among the criteria to identify IGD, with problematic players expected to engage in gaming activity for 30+ hours a week over the past 12 months \cite{b1}. None of the participants reported a weekly number of hours above the threshold (\textit{0-3}, n = 1, 9.09\%; \textit{4-10}, n = 4, 36.36\%; \textit{11-20}, n = 3, 27.27\%; \textit{21-30}, n = 3, 27.27\%, \textit{30+}, n = 0, 0\%).
The experimental session followed a structured sequence of gameplay and loot box interactions, ensuring that all participants experienced the same number of events. The session began with an exploration phase that was used as the baseline event to compare gameplay differences, during which players collected resources to unlock the first loot box interaction in the base. Upon returning to the base, they interacted with the loot box mechanic, which was presented with an animation and then followed by the presentation of the rewards, which consisted of weapon components, ordered by rarity (in ascending order, Rare, Valuable, and Unique). After interacting with the loot box, the players had to assemble the new weapon if they acquired the necessary components and then start the new exploration phase to find more resources. This sequence was repeated across three exploration and loot box phases, with each loot box interaction always allowing to obtain a new weapon. Each pair of loot box interaction and the following gameplay part is referred to as an in-game "Day". The final gameplay phase required participants to survive for two minutes, concluding the session when they returned to the base.

\subsection{Features Extraction}
Similarly to~\cite{b24}, EDA signals were filtered with a low-pass Butterworth filter with 0.159 Hz to remove high-frequency noise, and a high-pass Butterworth filter set to 0.05 Hz was used to correct baseline drifts. Data was synchronised with game markers, labelled (gP for the exploration phase and lB for the loot box interactions), and segmented. cvxEDA algorithm’s convex optimisation approach was implemented to extract the phasic and tonic components \cite{b27}. The phasic driver signal was used to compute SCRs by peak detection and AUC, calculated with a trapezoidal integral of the SCR over the epochs’ duration. The amplitude data for both SCR and SCL, both represented in microSiemens (µS) and were normalised using \eqref{eq1}.

\begin{equation}
\textit{SCR}=\frac{{\rm \textit{SCR}}_i}{{\rm \textit{SCR}}_{max}}\label{eq1}
\end{equation}

\section{Results}
\subsection{Arousal differences between gameplay and loot box interactions}
\begin{figure}[htbp]
\centerline{\includegraphics[scale=0.34]{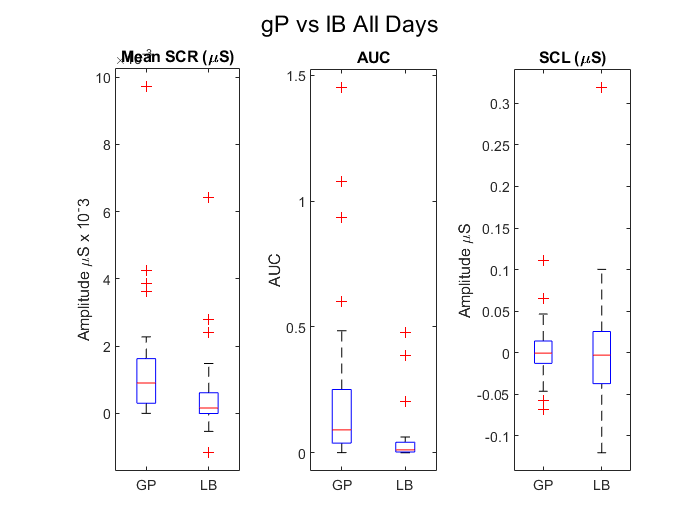}}
\caption{Comparison of gP and lB mean metrics across all days.}
\label{fig4}
\end{figure}

To address the first hypothesis about significant differences in arousal between gP and lB events, we divided the hypothesis testing into three parts, grouping game events with different configurations to examine from different perspectives how the game impacted physiological responses.
In the first part, we analyse metrics by grouping gP events 
and comparing them with the grouped lB events across all in-game Days,
with a total of 33 samples per group, to assess differences in the two game phases. Using the Shapiro-Wilk test, most of the analysed physiological metrics did not follow a normal distribution, necessitating the use of the nonparametric Mann-Whitney U test for statistical comparisons with a Bonferroni correction of 3, lowering the default alpha to .0167. In Table \ref{tab1}, the results of the Mann-Whitney U test show that the mean SCRs and AUC present a significant statistical difference when testing the two groups across all Days, with gP events eliciting a stronger skin response compared to the lB ones as shown in Fig. \ref{fig4}. In the second part of the analysis, the same test was performed on each Day's set of events (lB and following gP event) to evaluate specific trends. The statistical difference in mean SCRs and AUC is present on Day 1, but is excluded because of the Bonferroni correction, and in mean SCRs and AUC on Day 2. SCL did not show any statistical differences in all the tested combinations of events.

\begin{table}[h]
    \caption{Mann-Whitney U Test Results}
    \begin{center}
    \begin{tabular}{|l|cc| cc| cc|}
        \hline
        & \multicolumn{2}{c|}{\textbf{SCR}} & \multicolumn{2}{c|}{\textbf{AUC}} & \multicolumn{2}{c|}{\textbf{SCL}} \\
        & $p$ & $r$ & $p$ & $r$ & $p$ & $r$ \\
        \hline
        \textbf{All Days} & \textbf{.0009} & \textbf{.5782} & \textbf{.0002} & \textbf{.6496} & .6351 & .0826 \\
        \hline
        \textbf{Day0} & .3579 & .2772 & .115 & .4752 & 1 & 0 \\
        \textbf{Day1} & \textit{.0256} & \textit{.6732} & \textit{.0418} & \textit{.6138} & .5075 & .2915 \\
        \textbf{Day2} & \textbf{.0151} & \textbf{.7325} & \textbf{.0071} & \textbf{.8117} & .3676 & .3944 \\
        \hline
        \multicolumn{4}{l}{In \textbf{bold} the metrics below the alpha}
    \end{tabular}
    \label{tab1}
    \end{center}
\end{table}

\begin{figure}[htbp]
\centerline{\includegraphics[scale=0.34]{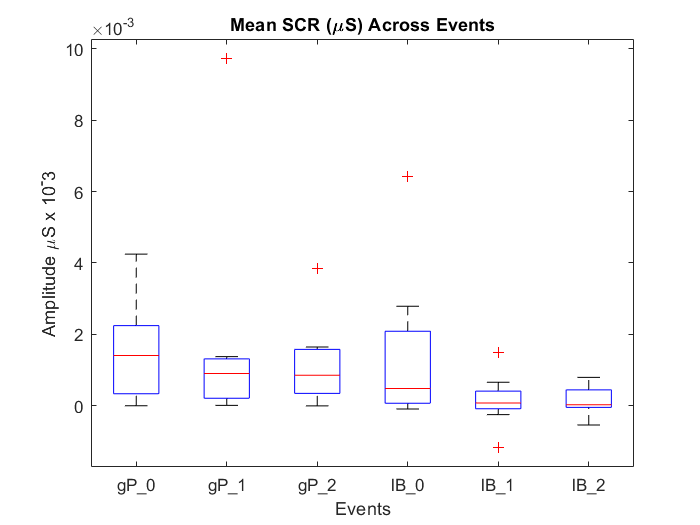}}
\caption{Mean SCR amplitude across gP and lB events.}
\label{fig5}
\end{figure}

\begin{figure}[htbp]
\centerline{\includegraphics[scale=0.34]{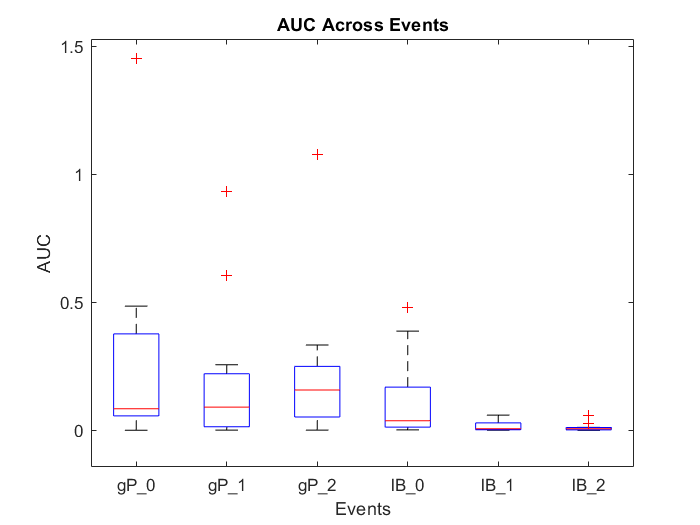}}
\caption{Mean AUC across gP and lB events.}
\label{fig6}
\end{figure}

For the final part of the analysis of the game events, in Table \ref{tab2}, we use the Friedman test to verify whether there are differences between the same types of events during the same session and identify trends. The results show that across gP events, none of the metrics present a statistical difference, while across lB events, the mean SCRs and AUC present a statistical difference. In Fig. \ref{fig5} and Fig. \ref{fig6}, mean SCR amplitudes and AUC respectively tend to keep similar values across the gP events, while it is noticeable how both metrics tend to decrease and flatten in lB events while progressing through the session with the peak registered for both measurements only on the first interaction during lB\_0.

\begin{table}[h]
    \centering
    \caption{Friedman Test Results}
    \begin{tabular}{|l|c|c|c|}
        \hline
        & \textbf{SCR \textit{p}-value} & \textbf{AUC \textit{p}-value} & \textbf{SCL \textit{p}-value} \\
        \hline
        \textbf{gP} & .1482 & .3067 & .4993 \\
        \hline
        \textbf{lB} & \textbf{.0116} & \textbf{.0039} & .1778 \\
        \hline
        \multicolumn{4}{l}{In \textbf{bold} the metrics below the alpha}
    \end{tabular}
    \label{tab2}
\end{table}


\subsection{Correlation with IGDS9-SF and Game Time}
SCRs and AUC were tested using the ranked Spearman correlation test, and results show that none of the parameters presents a statistically significant correlation when tested with the IGDS9-SF scores (\textit{SCRs}, \textit{p} = .88, \textit{r} = -.05; \textit{AUC}: \textit{p} = .42, \textit{r} = -.27) or the weekly hours spent gaming (\textit{SCRs}, \textit{p} = .87, \textit{r} = .06; \textit{AUC}: \textit{p} = .99, \textit{r} = .00). To test the hyposensitivity hypothesis based on~\cite{b24} results, the same test was executed between the SCRs ranges and both the IGDS9-SF scores and the weekly hours spent gaming. When testing this correlation, we tested the SCR amplitude ranges across all events (\textit{IGDS9-SF}: \textit{p} = .20, \textit{r} = -.41; \textit{Playtime}: \textit{p} = .75, \textit{r} = -.11), across gP events (\textit{IGDS9-SF}: \textit{p} = .10, \textit{r} = -.52; \textit{Playtime}: \textit{p} = .51, \textit{r} = -.22), and across lB events (\textit{IGDS9-SF}: \textit{p} = .99, \textit{r} = .00; \textit{Playtime}: \textit{p} = .55, \textit{r} = -.20). The results show that none of the tested ranges presents a statistically significant correlation with the IGDS9-SF scores or the weekly hours spent gaming. 

\section{Discussion and Conclusions}
This study examined the psychophysiological impact of loot box interactions in a single-player video game as a progression mechanic, revealing that core gameplay elicited stronger and more sustained arousal responses than randomised rewards, which showed declining effects over time, suggesting habituation or disengagement. No significant correlations emerged between arousal measures and IGDS9-SF scores or weekly game time, indicating that among non-problematic players, loot box engagement does not mirror the blunted arousal responses seen in related gambling research \cite{b24}. 
Our study focuses on loot boxes implemented as progression mechanics, as seen in video games like \textit{Balatro} and \textit{Vampire Survivors}, which lack social features, marketplaces, or the competitive component of other popular games that feature loot boxes. In multiplayer games like \textit{Counter-Strike}, these elements may elicit different physiological responses, suggesting that our findings may not fully generalize to such contexts. Moreover, the lack of participants with IGD scores above the high severity threshold restricted the examination of differences between problematic and non-problematic gamers. Future research should investigate the role of the outcome type and apply the proposed methodology to popular multiplayer games. However, this study represents an initial step in integrating physiological data into the broader conversation around problematic gambling and gaming behaviours and loot boxes, and contributes to a growing effort to establish psychophysiological methods as a valid approach to evaluate the player experience.

\bibliographystyle{IEEEtran}
\bibliographystyle{IEEEabrv, Thesis.bib}

\end{document}